\documentclass[showpacs,aps,twocolumn]{revtex4}

\usepackage{float}
\usepackage{bm}
\usepackage{amsmath}
\usepackage{graphicx}
\usepackage{subfigure}
\usepackage[usenames,dvipsnames]{color}
\definecolor{darkblue}{RGB}{0,0,196}
\usepackage[colorlinks=true,linkcolor=darkblue,citecolor=darkblue,urlcolor=darkblue]{hyperref}

\usepackage{setspace}
\usepackage{footmisc}
\usepackage[makeroom]{cancel}
\usepackage{comment}
\usepackage{lineno}
\usepackage{bbm}

\def\be{\begin{equation}}
\def\ee{\end{equation}}
\def\ba{\begin{eqnarray}}
\def\ea{\end{eqnarray}}

\usepackage{graphicx}
\usepackage{amsmath,bbm}
\usepackage{amssymb,bm}
\usepackage{yfonts}

\begin{document}
\title{Violation of Wiedemann-Franz Law for Hot Hadronic Matter created at NICA, FAIR and RHIC Energies using Non-extensive Statistics}
\author{Rutuparna Rath}
\author{Sushanta Tripathy}
\author{Bhaswar Chatterjee}
\author{Raghunath Sahoo}
\email{Raghunath.Sahoo@cern.ch}
\affiliation{Discipline of Physics, School of Basic Sciences, Indian Institute of Technology Indore, Simrol, Indore- 453552, INDIA}
\author{Swatantra Kumar Tiwari}
\affiliation{Department of Applied Science and Humanities, Muzaffarpur Institute of Technology, Muzaffarpur- 842003, Bihar}
\author{Abhishek Nath}
\affiliation{Department of Physical Sciences, Indian Institute of Science Education and Research,
Kolkata-741246, India}

\begin{abstract}
\noindent
We present here the computation of electrical and thermal conductivity by solving the Boltzmann transport equation in relaxation time approximation. We use the $q$-generalized Boltzmann distribution function to incorporate the effects of non-extensivity. The behaviour of these quantities with changing temperature and baryochemical potential has been studied as the system slowly moves towards thermodynamic equilibrium. We have estimated the Lorenz number at NICA, FAIR and the top RHIC energies and studied as a function of temperature, baryochemical potential and the non-extensive parameter, $q$. We have observed that Wiedemann-Franz law is violated for a non-extensive hadronic phase as well as for an equilibrated hadron gas at high temperatures.

\pacs{12.38.Mh, 24.10.Pa, 24.10.Nz, 25.75.-q, 47.75.+f}

\end{abstract}
\date{\today}
\maketitle 
\section{Introduction}
\label{intro}

The ultra-relativistic heavy-ion collision experiments like Relativistic Heavy Ion Collider (RHIC) at BNL and the Large Hadron Collider (LHC) at CERN provide a brief opportunity to look into a strongly interacting hot and dense matter consisting of deconfined quarks and gluons at an extreme temperature and/or energy density. The thermalized system of deconfined quarks and gluons, known as quark-gluon plasma (QGP) undergoes a phase transition to a system with hadronic degrees of freedom, which is known as the hadron gas (HG) phase as the system cools down. Large elliptic flow observed at RHIC hints that QGP is a strongly coupled near perfect fluid which necessitates a small but finite shear viscosity to entropy density ratio ($\frac{\eta}{s}$). The lowest bound of shear viscosity to entropy ratio ($\eta$/s) is 1/4$\pi$ as conjectured by AdS/CFT correspondence known as Kovtun-Son-Starinet (KSS) bound~\cite{Kovtun:2004de}. The measurements of elliptic flow in experiments like RHIC and LHC suggest that the system formed by the heavy-ion collisions have $\eta$/s close to KSS bound~\cite{Csernai:2006zz,Gyulassy:2004zy}. This signifies the importance of studying the transport properties to understand the hydrodynamic evolution properly.

In addition to the coefficients of viscosity, elliptic flow can also be significantly affected if there is some external or internal source of anisotropy. It has been suggested that extremely strong magnetic field ($\sim m_\pi^2$) might get produced at high energy collisions depending on centrality. Initially the magnetic field was thought to decay rapidly after collision \cite{Kharzeev:2007jp}. However, it has been pointed out that the electric field induced by a rapidly decaying magnetic field would resist farther decay and satisfy diffusion equation \cite{Tuchin:2013ie}. One crucially important thing in such a scenario is the electrical conductivity ($\sigma_{el}$) of the medium. Additionally, $\sigma_{el}$ also plays an important role in low mass dilepton production and the hydrodynamic evolution. So a proper estimation of $\sigma_{el}$ would be beneficial for overall better understanding of heavy ion collision events. Various methods have been proposed to estimate $\sigma_{el}$ like chiral perturbation theory~\cite{Fukushima:2008xe}, the numerical simulation of Boltzmann equation~\cite{Greif:2014oia}, holography~\cite{Finazzo:2013efa}, transport models~\cite{Cassing:2013iz}, Dyson-Schwinger equation~\cite{Qin:2013aaa} etc. Most of these calculations are for QGP and to understand the phase transition from QGP to HG, it is very important to learn about the electrical conductivity in HG.

Another critically important but less explored aspect of heavy ion collision is heat conduction. At very high energy collision events, this may be of lesser 
importance as the net baryon density is very small. But at NICA and FAIR energies as well as at low energy runs at RHIC, baryon density can be significant and heat conduction can play a major role in the evolution of the system. In the hadronic phase, thermal conductivity ($\kappa$) has been calculated for pion gas by assigning a pion chemical potential as pion numbers can be taken to be constant in the late stage of the evolution \cite{Mitra:2014dia}. Recently it has also been calculated for baryonic matter within hadron resonance gas (HRG) model \cite{Kadam:2017iaz}.

To describe the particle production mechanism and study the QCD thermodynamics, the statistical models are more useful due to high multiplicities produced in high-energy collisions. It has been proposed that the transverse momentum ($p_{\rm{T}}$) spectra of final state particles produced in high-energy collisions would follow a thermalized Boltzmann-Gibbs (BG) distribution. However, a finite degree of deviation from the BG distribution of the $p_{\rm{T}}$-spectra has been observed by the RHIC~\cite{star-prc75,phenix-prc83} and LHC~\cite{alice1, alice2, alice3} experiments. Also, the matter produced in these extreme conditions evolves rapidly in non-homogenous way. Thus, the global equilibrium is not necessarily established and a power law tail develops in the particle spectra instead of exponential distributions. Recently, It has been observed that the particle spectra in high energy hadronic and heavy-ion collisions are successfully explained by the Tsallis non-extensive statistics ~\cite{Wilk:1999dr,Tsallis:1987eu,Tsallis:1999nq,Thakur:2016boy, Sett:2015lja, Bhattacharyya:2015hya, Zheng:2015gaa, Tang:2008ud, De:2014dna}, which is purely motivated by the spectral shape of the identified particles.
Using Tsallis non-extensive statistics in Boltzmann transport equation (BTE) with relaxation time approximation (RTA) has successfully explained the elliptic flow~\cite{Tripathy:2017nmo} and nuclear modification factor~\cite{Tripathy:2016hlg, Tripathy:2017kwb} of identified particles. The dissipative properties such as shear and bulk viscosity has been reported using the non-extensive Boltzmann Transport Equation (NBTE) in RTA~\cite{Kakati:2017xvr}.

In this work, we study the behaviour of $\sigma_{el}$ and $\kappa$ with temperature at finite baryon chemical potential, $\mu_B$ in non-extensive scenario. We use the non-extensive Boltzmann distribution function and solve the BTE  using RTA to obtain $\sigma_{el}$ and $\kappa$ and study their behaviour for different values of the non-extensive parameter. We also calculate the ratio of the two coefficients to examine the validity of Wiedemann-Franz law which was originally proposed for free electron metals. 

We have organized the paper in the following manner. In the Sect.~\ref{formulation}, we derive electrical and thermal conductivities for hadronic degrees of freedom using BTE in RTA. Here, we take $q$-equilibrium as a solution of BTE. Section~\ref{result} presents the discussion of results obtained using the formulation. In Sect.~\ref{summary}, the summary and conclusions of this work are presented.

\section{Formulation}
\label{formulation}
\subsection{Electrical Conductivity}
\label{Electrical Conductivity}
The transport properties such as electrical conductivity and thermal conductivity for a hadronic matter using non-extensive statistics has been calculated by the technique mentioned in Ref.~\cite{ref1,Hosoya:1983xm}. From the Relativistic Boltzmann Transport equation, we can initiate our derivation of transport coefficients which is given by,
\begin{equation}
p^{\mu}\partial_{\mu}f_a(x , p) + Q_aF^{\alpha\beta}p_{\beta}\frac{\partial f_a(x , p)}{\partial p^\alpha} = C_a[f_a],
\label{eq1}
\end{equation}
where $p^{\mu}$ is the momentum four vector, $Q_a$ is the electric charge of the $a^{th}$ particle, $f_a(x , p)$ is the particle distribution function when system is away from equilibrium, index $a$ is used in the distribution function for different hadronic species and $F^{\alpha\beta}$ is the electromagnetic field strength tensor, which is given by,
\[ \left( \begin{array}{cccc}
0 & -E_{x} & -E_{y}  & -E_{z} \\
E_{x} & 0 & 0  & 0 \\
E_{y} & 0 & 0  & 0 \\
E_{z} & 0 & 0  & 0 \end{array} \right)\]
Here we have chosen a frame where there is no magnetic field.
$C_a[f_a]$ is the collision integral term which is approximated using the relaxation-time approximation (RTA) and given by
 \begin{equation}
C_a[f_a] \simeq -\frac{p^{\mu}u_{\mu}}{\tau_a}\delta f_a,
\label{eq2}
\end{equation}
where $u_{\mu}$ = (1,  $\bf 0$) is the fluid four velocity in the local rest frame and $\tau_a$ is the relaxation time of the system, which is the time required by the system to reach the $q$-equilibrium. 

Considering the system being relaxed towards the equilibrium state, we can take the function in the form:
 \begin{equation}
f_a(x , p) = f_a^0(x , p) + \delta f_a = f_a^0(x , p) [1+ \phi(x , p)],
\label{eq3}
\end{equation}
where $\phi (|  \phi | << 1)$ is used for perturbation. We have taken the non-extensive Tsallis distribution as $f_a^0$~\cite{Bezerra:2002gi} near the local rest frame of the fluid, where the system is described locally by temperature, $T$, baryochemical potential, $\mu_B$ and fluid velocity, $u_\mu$, which change slowly in space and time~\cite{Gavin:1985ph}. In Boltzmann's approximation, the thermodynamically consistent Tsallis distribution is given as,
\begin{equation}
f_a^0 = \frac{1}{\Big[1 + (q-1)\Big(\displaystyle \frac{{p^{\mu}u_{\mu} } - \mu}{T}\Big)\Big]^{\displaystyle \frac{q}{q-1}}}.
\label{eq4}
\end{equation}
 $q$ is the non-extensive parameter, which signifies how far the system is away from thermodynamic equilibrium. $T$ and $\mu= B\mu_B+ s\mu_s$ are temperature and chemical potential, respectively. Here we consider only the baryochemical potential, $\mu_B$, ignoring the strangeness chemical potential. By  applying local equilibrium approximation ($f_a \equiv f_a^0$) in the LHS of BTE, we obtain:
\begin{equation}
Q_a\Big(p_0 {\bf E} . \frac {\partial f_a^0}{\partial {\bf p}} + {\bf E} .{\bf p} \frac{\partial f_a^0}{\partial p_0} \Big)= -\frac{p_{0}}{\tau_a}\delta f_a.
\label{eq5}
\end{equation}
Now, for a constant electric field $\bf E$, Eq. (5) becomes 
\begin{equation}
\delta f_a = q\frac{Q_a\tau_a{\bf E}.{\bf p}}{T p_{0}} f_0^{(2q-1)/q}.
\label{eq6}
\end{equation}
 As we know, the electrical conductivity is a parameter which quantifies the response of the system to an applied electric field. Using the relationship between electric current ({\bf j}) and applied electric field ({\bf E}), one can have the expression for $\sigma_{el}$ as,
 \begin{equation}
{\bf j} = \sigma_{el} {\bf E}.
\label{eq7}
\end{equation}
The four current, $j^{\mu} = (j, {\bf j})$ is defined as,
\begin{equation}
j^{\mu}= Q_{a} g_{a}\int \frac{d^3p}{(2\pi)^3 E_{a}} p^{\mu}f_{a}(x , p),
\label{eq8}
\end{equation}
where $g_a$ is the degeneracy and $E_a^2 = p^2 +m_a^2$ is the energy dispersion relation for the $a^{th}$ hadron species.
 When we are slightly away from the equilibrium, we can apply the approximation mentioned in Eq. \ref{eq3} to $j^{\mu}$ and obtain the four current as:
 \begin{equation}
 j^{\mu} = j_{0}^{\mu} + \triangle j^{\mu}.
\label{eq9}
\end{equation}
So, \begin{equation}
\begin{split}
 j^{\mu} &= Q_{a} g_{a}\int \frac{d^3p}{(2\pi)^3 E_{a}} p^{\mu}f_{a} + Q_{a} g_{a}\int \frac{d^3p}{(2\pi)^3 E_{a}} p^{\mu}\delta f_{a} \\
 & = j_{0}^{\mu} + \triangle j^{\mu},
\label{eq10}
\end{split}
\end{equation}
and \begin{equation}
 \triangle j^{\mu} = Q_{a} g_{a}\int \frac{d^3p}{(2\pi)^3 E_{a}} p^{\mu}\delta f_{a},
\label{eq11}
\end{equation}
Using local equilibrium approximation, we can write
 \begin{equation}
{\bf j} = {\bf \triangle j}=\sigma_{el} {\bf E}.
\label{eq12}
\end{equation}
Using Eq.~\ref{eq6}, \ref{eq11} and \ref{eq12} the electrical conductivity ($\sigma_{el}$) can be evaluated. We can start with $E_{x}$ component:
 \begin{equation}
 \begin{split}
 \triangle j^{x} & = Q^{2}_{a} g_{a}\int \frac{d^3p \tau_a}{T(2\pi)^3 E^{2}_{a}} p_{x}(E_{x}p_{x}+E_{y}p_{y}\\
 &+E_{z}p_{z}) q f_0^{(2q-1)/q}\\
 & = \sigma_{xx}E_{x}+\sigma_{xy}E_{y}+\sigma_{xz}E_{z}.
\end{split}
\label{eq13}
\end{equation}
Since electrical conductivity is related to the $E_{x}$ part of the above equation, we can define (using the fact that $p_{x}^{2}= p_{y}^{2} = p_{z}^{2} = \frac{\bf p^{2}}{3}$)
\begin{equation}
\sigma_{xx}E_{x}= \sigma_{el} E_{x} = Q^{2}_{a} g_{a}\int \frac{d^3p{p^{2}}{\tau_a}}{3T(2\pi)^3 E^{2}_{a}} q f_0^{(2q-1)/q}E_{x}.
\label{eq14}
\end{equation}
Hence,\begin{equation}
\sigma_{el}= \frac{1}{3T}\sum_a Q^{2}_{a} g_{a}\int \frac{d^3p{p^{2}}{\tau_a}}{(2\pi)^3 E^{2}_{a}} q f_0^{(2q-1)/q}.
\label{sigmael}
\end{equation}
\vspace*{0.1cm}
\subsection{Thermal Conductivity}
\label{Thermal Conductivity}
The heat flow in the interacting systems can be described by the quantity called thermal conductivity $\kappa$. The energy momentum tensor $T^{\mu\nu}$ and the four current $j^{\mu}$ is given by;

\begin{equation}
T^{\mu\nu}= \ g_{a}\int {\frac{d^3p{p^{\mu}{p^{\nu}}}}{(2\pi)^3 E_{a}}}{f_a(x, p)},
\label{eq15}
\end{equation}

\begin{equation}
j^{\mu}= \ g_{a}\int {\frac{d^3p{p^{\mu}}}{(2\pi)^3 E_{a}}}{f_a(x, p)}.
\label{eq16}
\end{equation}
For a small perturbation from the equilibrium, the change in the energy momentum tensor $\triangle T^{\mu\nu}$ and the four current $\triangle j^{\mu}$ can be written as,

\begin{equation}
\triangle T^{\mu\nu}= \ g_{a}\int {\frac{d^3p{p^{\mu}{p^{\nu}}}}{(2\pi)^3 E_{a}}}{\delta f_a},
\label{eq17}
\end{equation}

\begin{equation}
\triangle j^{\mu}= \ g_{a}\int {\frac{d^3p{p^{\mu}}}{(2\pi)^3 E_{a}}}{\delta f_a}.
\label{eq18}
\end{equation}

The $\delta f_a$ term can be evaluated from the collision term present in the above BTE in the absence of external field. Hence one can write,

 \begin{equation}
p^{\mu}\partial_{\mu} f_{a}(x , p)=-\frac{p^{\mu}u_{\mu}}{\tau_a}\delta f_a,
\label{eq19}
\end{equation}
 where $\partial_{\mu} = u_{\mu}D + \bigtriangledown_{\mu}$ and the convective derivatives $(DT,D_{\mu},Du^{\mu})$ can be decimated by using the following relations as done in \cite{Hosoya:1983xm} ;
\begin{equation}
(\epsilon+P)Du^{\mu}-\bigtriangledown^{\mu}P=0,
\label{eq20}
\end{equation}
\begin{equation}
Dn+n\bigtriangledown_{\mu}u^{\mu}=0.
\label{eq21}
\end{equation}
Using the above two relations, $\triangle T^{\mu\nu}$ and $\triangle j^{\mu}$ can be written as,
\begin{equation}
\begin{split}
\triangle T^{\mu\nu}= &\ g_{a}\int {\frac{d^3p}{(2\pi)^3 E_{a}}} {\frac{{p^{\mu}}{p^{\nu}}}{p.u}}{\frac {1}{T}}\Bigg[ \tau_{a}q f_0^{(2q-1)/q}\bigg\{p.u{(\frac{\partial P}{\partial \varepsilon})}_{n}\\
&\nabla_{\alpha}u^{\alpha}+p^{\alpha}X_{\alpha}+\frac{p^{\alpha}p^{\beta}}{p.u} \nabla_{\alpha}u_{\beta}+(\frac{\partial P}{\partial n})_{\varepsilon}\nabla_{\alpha}u^{\alpha}\\
&-\frac{\varepsilon+P}{n}\frac{p^{\alpha}}{p.u}X_{\alpha}\bigg\}\Bigg],
\end{split}
\label{eq22}
\end{equation}
\begin{equation}
\begin{split}
\triangle j^{\mu}= &\ g_{a}\int {\frac{d^3p}{(2\pi)^3 E_{a}}}{\frac{{p^{\mu}}}{p.u}}{\frac {1}{T}}\Bigg[ \tau_{a}q f_0^{(2q-1)/q}\bigg\{p.u{(\frac{\partial P}{\partial \varepsilon})}_{n}\\
&\nabla_{\alpha}u^{\alpha}+p^{\alpha}X_{\alpha}+\frac{p^{\alpha}p^{\beta}}{p.u} \nabla_{\alpha}u_{\beta}+(\frac{\partial P}{\partial n})_{\varepsilon}\nabla_{\alpha}u^{\alpha}\\
&-\frac{\varepsilon+P}{n}\frac{p^{\alpha}}{p.u}X_{\alpha}\bigg\}\Bigg],
\end{split}
\label{eq23}
\end{equation}
where
\begin{equation}
X_{\alpha}=\frac{\nabla_{\alpha}P}{\varepsilon+P}-\frac{\nabla_{\alpha}T}{T},
\label{eq24}
\end{equation}
and $u_{\mu}=(1,\bf0)$. $\varepsilon$ and $n$ are the energy density and number density respectively. $ T^{0i}=\Delta T^{0i}-\frac{(\varepsilon+P)}{n}\Delta j^{i}\equiv I^{i} $.
\begin{equation}
\Delta T^{0i}=\sum_{a}g_{a}\int \frac{d^3p}{(2\pi)^3}{\frac{ \bf{p^{2}}}{3T}}\tau_{a}q f_0^{(2q-1)/q}\bigg\{1-\frac{\varepsilon+P}{nE_a}\bigg\}X_{i}
\label{eq25}
\end{equation}
 and
 \begin{equation}
\Delta j^{i}=\sum_{a}g_{a}\int \frac{d^3p}{(2\pi)^3E_{a}}\frac{ \bf{p^2}}{3T}\tau_{a}q f_0^{(2q-1)/q}\bigg\{1-\frac{\varepsilon+P}{nE_a}\bigg\}X_{i}.
\label{eq26}
\end{equation}
Using the Eckart condition, heat conductivity can be defined as,
\begin{equation}
I^{i}=-\kappa\left[\partial_{i}T-T\partial_{i}P/(\varepsilon+P)\right]=\kappa T X_{i}.
\label{eq27}
\end{equation}  
Now the thermal conductivity is defined as,
\begin{equation}
\kappa = \frac{1}{3T^{2}}\sum_{a}g_{a}\tau_{a}\int \frac{d^3p}{(2\pi)^3}\frac{\bf p^2}{E_{a}^2}q f_0^{(2q-1)/q}\left(E_{a}-\frac{t_a\omega}{n}\right)^{2},
\label{kappa}
\end{equation}
where $ \omega=\varepsilon+P $ is the enthalpy and $ t_a=+1(-1) $ for particles (anti-particles).
\subsection{Relaxation Time}
\label{Relaxation Time}

The energy dependent relaxation time is given as, 
\begin{equation}
\tau^{-1} (E_a) = \sum_{bcd} \int \frac{d^3p_b d^3p_c d^3p_d}{(2\pi)^3 (2\pi)^3 (2\pi)^3} W(a,b \rightarrow c,d) f_b^0,
\label{eq29}
\end{equation}
where the transition rate $W(a,b \rightarrow c,d)$ is defined as,
\begin{equation}
W(a,b \rightarrow c,d) = \frac{{2\pi}^4\delta(p_a+p_b-p_c-p_d)}{2E_a2E_b2E_c2E_d} |\mathcal{M}|^2,
\label{eq30}
\end{equation}
and the transition amplitude is $|\mathcal{M}|$. By considering the center-of-mass frame, Eq.~\ref{eq29} can be simplified as,
\begin{equation}
\begin{split}
\tau^{-1}(E_a) &= \sum_{b} \int \frac{d^3p_b}{(2\pi)^3} \sigma_{ab} \frac{\sqrt{s - 4m^2}}{2E_a2E_b} f_b^0 \\
&\equiv \sum_{b} \int \frac{d^3p_b}{(2\pi)^3} \sigma_{ab} v_{ab} f_b^0,
\end{split}
\label{eq31}
\end{equation}
where, $v_{ab}$ and $\sqrt{s}$ are the relative velocity and the center-of-mass energy, respectively. The total scattering cross-section in the process $a(p_a) + b(p_b) \rightarrow a(p_c) + b(p_d)$ is given by $\sigma_{ab}$. $\tau(E_a)$ can be approximated to averaged relaxation time ($\widetilde{\tau}$)~\cite{Moroz:2013haa} for further simplification and it can be done by averaging over $f_a^0$ using Eq.~\ref{eq31} and given as,
%\textcolor{blue}{
\begin{equation}
\begin{split}
\widetilde{\tau_a}^{-1} &= \frac{\int \displaystyle \frac{d^3p_a}{(2\pi)^3}\tau^{-1}(E_a) f_a^0}{\int \displaystyle \frac{d^3p_a}{(2\pi)^3}f_a^0}\\
& = \sum_{b} \displaystyle \frac{\int \displaystyle \frac{d^3p_a}{(2\pi)^3} \displaystyle \frac{d^3p_b}{(2\pi)^3} \sigma_{ab}v_{ab}f_a^0f_b^0}{\int \displaystyle \frac{d^3p_a}{(2\pi)^3}f_a^0} \\
 &= \sum_{b} n_b \langle \sigma_{ab} v_{ab}\rangle,
\end{split}
\label{eq32}
\end{equation}
%\vspace*{0.5cm}
here, $ n_b = \int \frac{d^3p}{(2\pi)^3}f_b^0$ is the number density of $b^{th}$ hadronic species. As derived in~\cite{Kakati:2017xvr}, the thermal average for the scattering of same species of particles at a given $T$ and $\mu_{B}$ with constant cross-section is given as,

\begin{widetext}
\begin{equation}
\left\langle \sigma_{ab}v_{ab} \right \rangle =  \frac{ \sigma \int 8\pi^2 p_a p_b dE_a dE_b d\cos\theta~e_{q}^{-E_{a}/T}e_{q}^{-E_{b}/T}\times \frac{\sqrt{(E_aE_b-p_ap_b\cos\theta)^2-(m_am_b)^2}}{E_aE_b-p_ap_b\cos\theta}}{\int 8\pi^2 p_a p_b dE_a dE_b d\cos\theta~e_{q}^{-E_{a}/T}e_{q}^{-E_{b}/T}}.
\label{sigvab}
\end{equation}
\end{widetext}

Here, $e_{q}^x$ is the q-exponential which is defined as  $e_{q}^x = \left[ 1+(q-1)x\right]^{q/(q-1)}$. The cross-section $\sigma$ is used as a parameter in the calculations. $E_a$ and $E_b$ are integrated in the limit $m_a$ to $\infty$ and $m_b$ to $\infty$, respectively. The integration limit for $\cos\theta$ is -1 to 1. The other thermodynamical quantities using non-extensive statistics are calculated as~\cite{ref5},

\begin{alignat}{3}
& n = g \int \frac{d^3p}{(2\pi)^3} \left[1 + (q-1) \frac{E-\mu}{T}\right] ^ {-\frac{q}{q-1}}, \label{eq38} \\
& \epsilon = g \int \frac{d^3p}{(2\pi)^3} E \left[1 + (q-1)\frac{E-\mu}{T}\right] ^ {-\frac{q}{q-1}}, \label{eq39} \\
& P= g \int \frac{d^3p}{(2\pi)^3} \frac{p^2}{3E} \left[1 + (q-1)\frac{E-\mu}{T}\right] ^ {-\frac{q}{q-1}},
\label{eq40}
\end{alignat}
where $n$, $\epsilon$ and $P$ are the number density, energy density and pressure of hadrons, respectively.

\section{Results and Discussions}
\label{result}

\begin{figure}[h]
\includegraphics[height=22em]{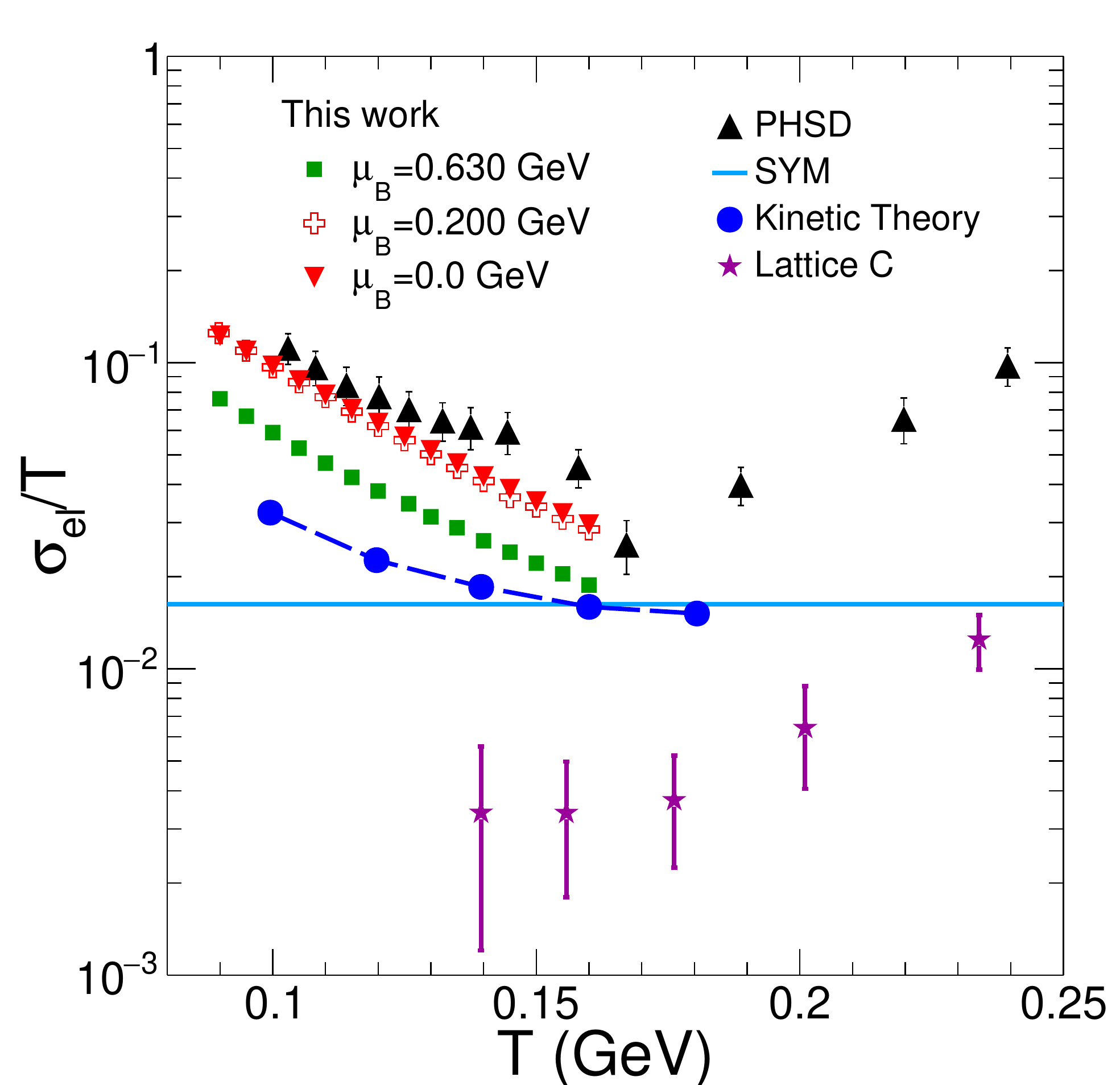}
\caption[]{(Color online) $\sigma_{el}/T$ vs $T$ for different baryochemical potentials, $\mu_B$ with  $q \sim$ 1 are compared with the results from other theoretical models.}
\label{fig1}
\end{figure}

In this section, we present our findings and corresponding analysis regarding the electrical and thermal conductivities which are obtained by solving the 
Boltzmann transport equation with non extensive statistics. The parameter $q$, which denotes how far from thermal equilibrium the system is, affects the relaxation time. By examining the electrical conductivity ($\sigma_{el}$) and the thermal conductivity ($\kappa$), we investigate how the system behaves with changing $q$-values for different temperature ($T$) and baryon chemical potential ($\mu_B$). For the studies done in this paper, we have chosen a temperature range of $T= 90-160$ MeV, as a possible hadronic phase is expected to be created below the critical temperature, $T_c \sim 160$ MeV. We have limited ourself to $q<1.15$ as it has been shown 
in Ref.\cite{Khuntia:2016ikm} that beyond that value there might not be any phase transition which would in effect render the study irrelevant. It is necessary to include all hadrons up to a certain cutoff. For our analysis, we have chosen the mass cutoff, $\Lambda=2.0$ GeV. For simplicity, we have considered constant cross-section, $\sigma$ = 200 \textit{mb} and a universal freeze-out scenario with same $q$-values for all hadron species. Though, for a more precise description, the temperature and chemical potential dependence of the cross-section should be considered.

First, we want to see the variation of temperature scaled- $\sigma_{el}$ and $\kappa$ (to make them dimensionless) with $q$ for different temperature and $\mu_B$ and then we shall examine the ratio $\frac{\kappa}{\sigma_{el}}$ to see if it is possible to find analogy of hadronic gas within condensed matter systems. For our analysis, we have varied the temperature in the range from 90 MeV to 160 MeV. We have chosen five different baryon chemical potentials: $\mu_{B} = 25$, $45$, $200$, $436$ and $630$ MeV, which is relevant for RHIC at $\sqrt{s_{NN}}=200$, $130$, $19.6$ GeV, RHIC/FAIR at $\sqrt{s_{NN}}=7.7$ GeV and NICA at $\sqrt{s_{NN}}=3$ GeV, respectively~\cite{Tawfik:2016sqd, BraunMunzinger:2001ip, Cleymans:2005xv, Khuntia:2018non}.

 %$\mu_B=25$ MeV which is relevant for RHIC at $\sqrt{s_{NN}}=200$ GeV, $\mu_B=200$ MeV relevant for RHIC at $\sqrt{s_{NN}}=19.6$ GeV, 
%$\mu_B=436$ MeV relevant for CMB at $\sqrt{s_{NN}}=7.7$ GeV and $\mu_B=630$ MeV relevant for NICA at $\sqrt{s_{NN}}=3$ GeV ~\cite{Tawfik:2016sqd, BraunMunzinger:2001ip, Cleymans:2005xv, Khuntia:2018non}.

 First, we want to see how our results compare with results obtained in previous literature. For that, in Fig. \ref{fig1}, we have shown our results for electrical conductivity along with results obtained using other approaches. For this, we have taken $q=1.0001$ so that it approximates the equilibrium scenario since results obtained in other approaches assume equilibrium distribution function. In Fig. \ref{fig1}, we have considered three different baryon chemical potentials, $\mu_B$. We see that $\sigma_{el}$ decreases with $\mu_B$ which we shall discuss in detail in following plots. We have compared our results with those obtained using lattice QCD \cite{Gupta:2003zh}, super Yang-Mills plasma \cite{CaronHuot:2006te}, kinetic theory \cite{Greif:2016skc} and parton hadron string dynamics (PHSD) \cite{Cassing:2013iz}. Lattice QCD data for $\sigma_{el}$/T for hot QCD plasma are extracted from a quenched lattice measurement of the Euclidean time vector correlator as shown in the figure by the star markers. The cyan horizontal line is the result obtained for super Young-Mills plasma. Blue circles are the kinetic theory results for hadrons which also decrease with temperature along with other results for hadronic matter. The results obtained in PHSD model are also shown by the black triangles in the figure for both the phases- hadron gas and quark-gluon plasma with different approaches. In PHSD, the hadronic sector is equivalent to the Hadron- String-Dynamics (HSD) transport approach which is a covariant extension of the Boltzmann-Uehling-Uhlenbeck (BUU) approach. We can see that the values we obtained are slightly less compared to the values obtained in PHSD and significantly higher compared to the estimations done using kinetic theory for $\mu_B=0$.

\begin{figure}[h]
\includegraphics[height=22em]{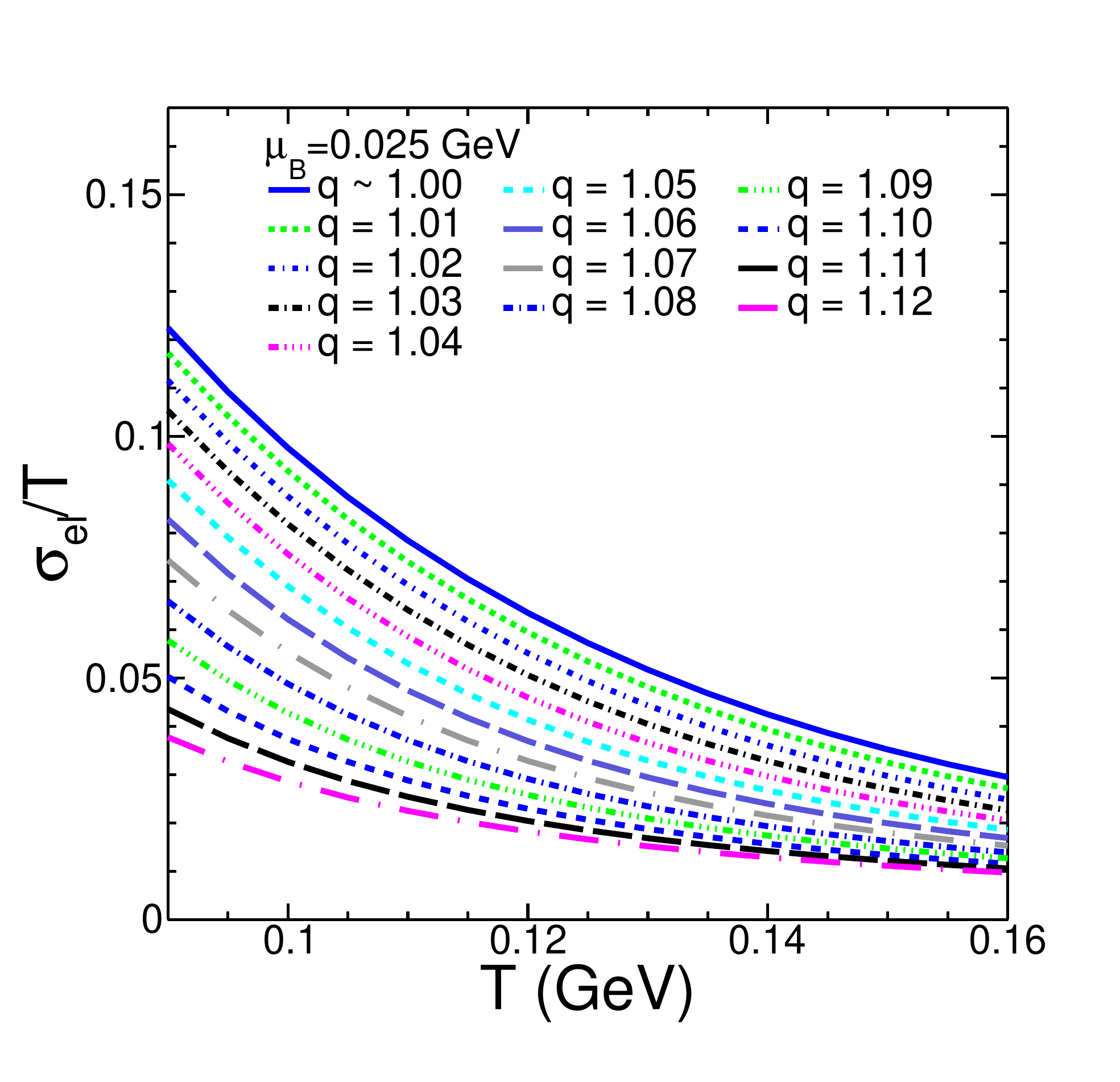}
\caption[]{(Color online) $\sigma_{el}/T$ vs $T$ for different $q$-values at $\mu_B=0.025$ GeV.}
\label{sigmael50}
\end{figure}

\begin{figure}[h]
\includegraphics[height=22em]{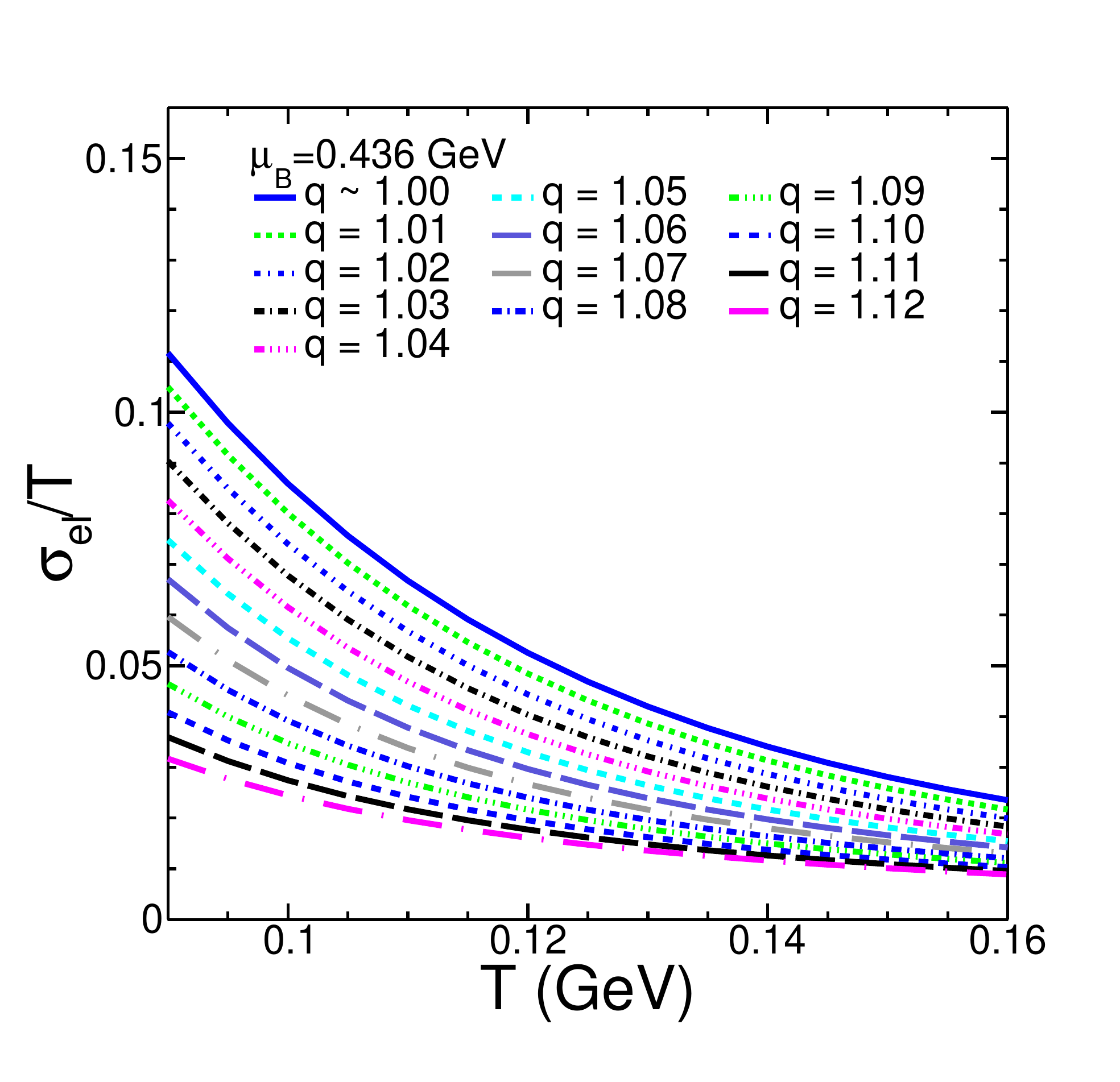}
\caption[]{(Color online) $\sigma_{el}/T$ vs $T$ for different $q$-values at $\mu_B=0.436$ GeV}
\label{sigmael100}
\end{figure}

In Fig.\ref{sigmael50} and Fig.\ref{sigmael100},  the temperature scaled-$\sigma_{el}$ is shown as a function of temperature for different $q$-values for $\mu_B=25$ MeV and 436 MeV. We can clearly see that $\sigma_{el}$ decreases with both temperature and the Tsallis parameter. $q=1.0001$ roughly corresponds to equilibrium Boltzmann distribution case and higher values of $q$ signifies 
moving farther away from equilibrium. At $q=1.0001$, our result shows similar behaviour as observed in Ref.\cite{Kadam:2017iaz} and the values are also in the same ballpark. We observe that 
$\sigma_{el}$ decreases more rapidly at lower temperatures, whereas at higher temperature the fall is more gradual. Also, we see that the sharp decrease is more prominent at lower $q$ values. This can be understood by examining carefully the expression of $\sigma_{el}$ in Eq.[\ref{sigmael}]. Apart from the factor of $\frac{1}{T}$, there are two quantities in the expression which depends on $T$, $\mu_B$ and $q$: the relaxation time $\tau$ and the distribution function $f_0$. Of these two quantities, $\tau$ decreases sharply with temperature 
for lower temperature and falls gradually at higher temperature. On the other hand, $f_0$ increases with temperature but its increase is not sharp enough to combat the fall of $\tau$. So $\sigma_{el}$ roughly mimics the behaviour of $\tau$. With $q$ also, the same pattern follows. This behaviour of $\sigma_{el}$ can also be intuitively understood. At higher value of temperature or $q$, the system experiences more collision between hadrons which effectively reduces the flow of charged particles and hence $\sigma_{el}$ decreases.

We observe an apparently strange behaviour of $\sigma_{el}$ with $\mu_B$. From the two figures, it is clear that $\sigma_{el}$ barely changes with $\mu_B$ whereas we would expect it to decrease noticeably since $\tau$ decreases significantly. However this can also be understood by examining Eq.[\ref{sigmael}] and the expression for $\langle\sigma_{ab}v_{ab}\rangle$ in Eq.[\ref{sigvab}]. The dependence of  $\langle\sigma_{ab}v_{ab}\rangle$ on $\mu_B$ is very weak, resulting in the relaxation time $\tau$ to depend on $\mu_B$ in a similar fashion, which is almost exactly countered by the opposite behaviour of $f_0$ with $\mu_B$. Even more significantly, the contributions to $\sigma_{el}$ comes only from electrically charged hadrons which in our case is dominated by charged mesons, which does not depend explicitly on $\mu_B$. 

\begin{figure}[h]
\includegraphics[height=22em]{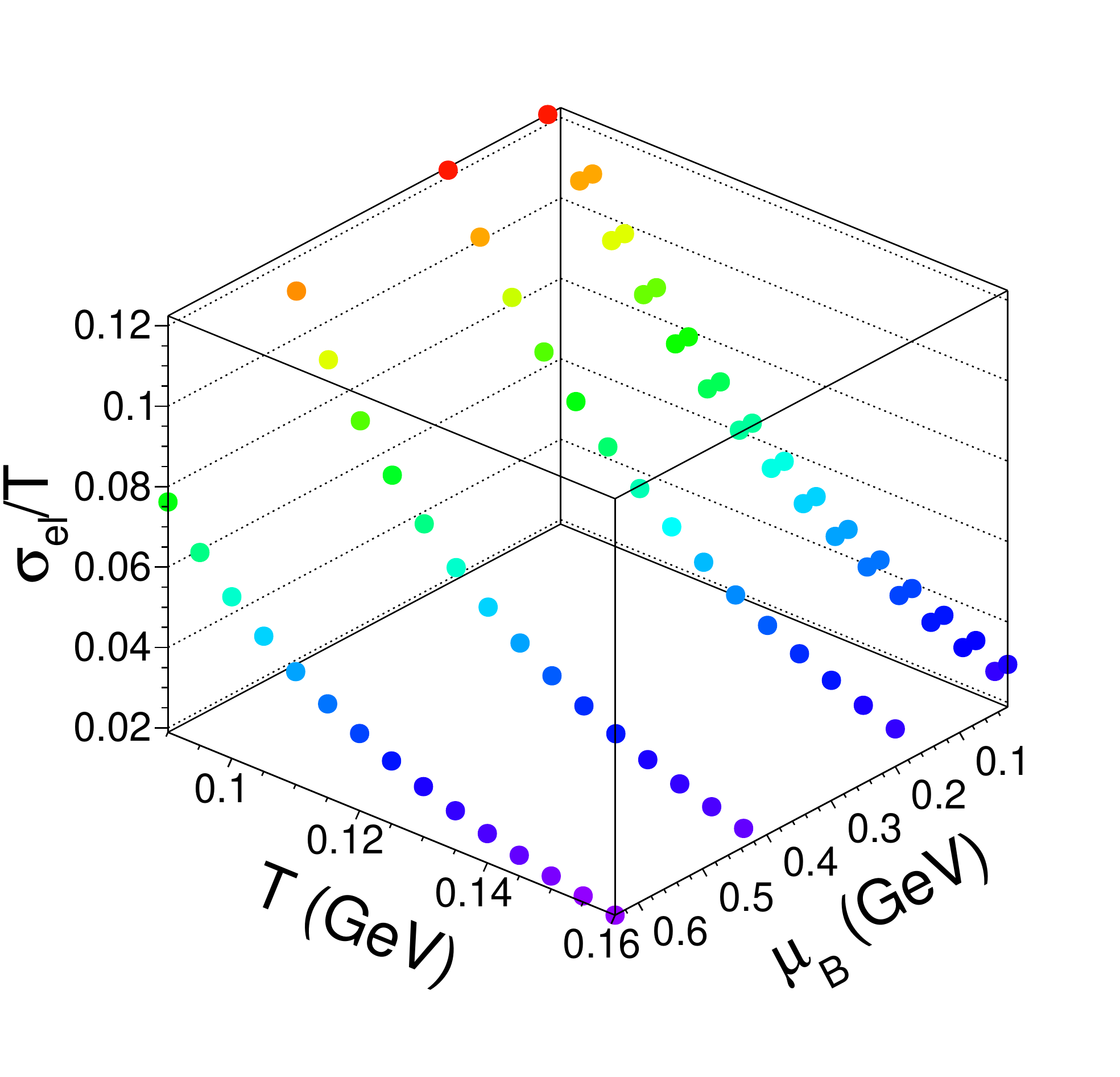}
\caption[]{(Color online)  $\sigma_{el}/T$  as a function of $T$ and $\mu_{B}$ for different center of mass energies at q $\sim 1$}
\label{sigma3D}
\end{figure}

Figure \ref{sigma3D} shows the variation of $\sigma_{el}$ both with $T$ and $\mu_B$ simultaneously for the equilibrium scenario. Here, we have included results for all different $\mu_B$ that has been considered. It is obvious that $\sigma_{el}$  remains more or less unchanged up to $\mu_B = 436$ MeV and beyond that, the change is significant at lower temperatures. At higher temperature, the change in $\sigma_{el}$ is negligible as it approaches extremely small value for all $\mu_B$. This is because at higher collision energies (low $\mu_B$) the produced system is meson (dominantly pions) dominated, whereas at lower collision energies the system is baryon rich.

\begin{figure}[h]
\includegraphics[height=22em]{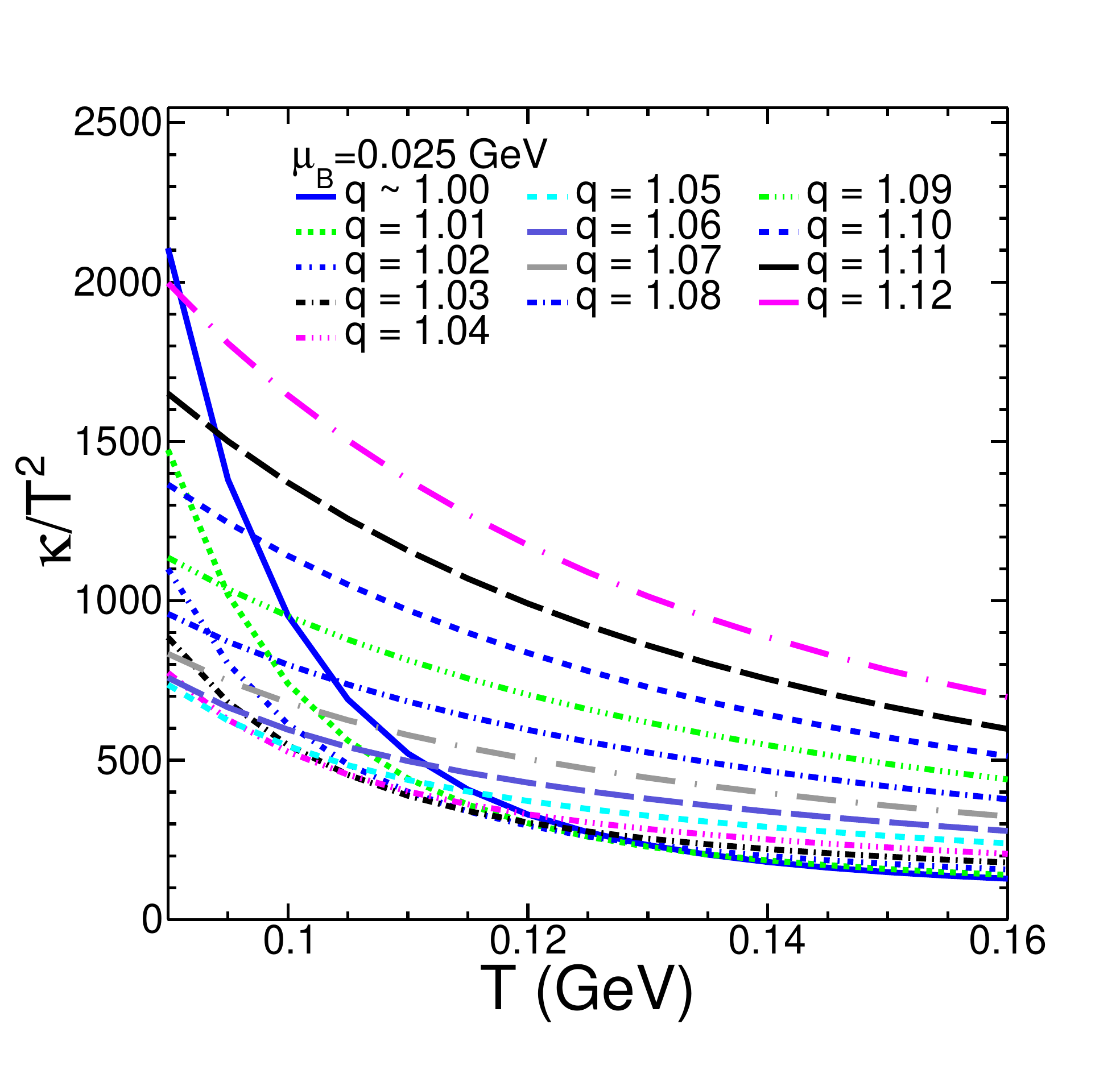}
\caption[]{(Color online) $\kappa/T^{2}$ vs $T$ plot for different $q$-values at $\mu_B=0.025$ GeV. }
\label{kappa25}
\end{figure}

\begin{figure}[h]
\includegraphics[height=22em]{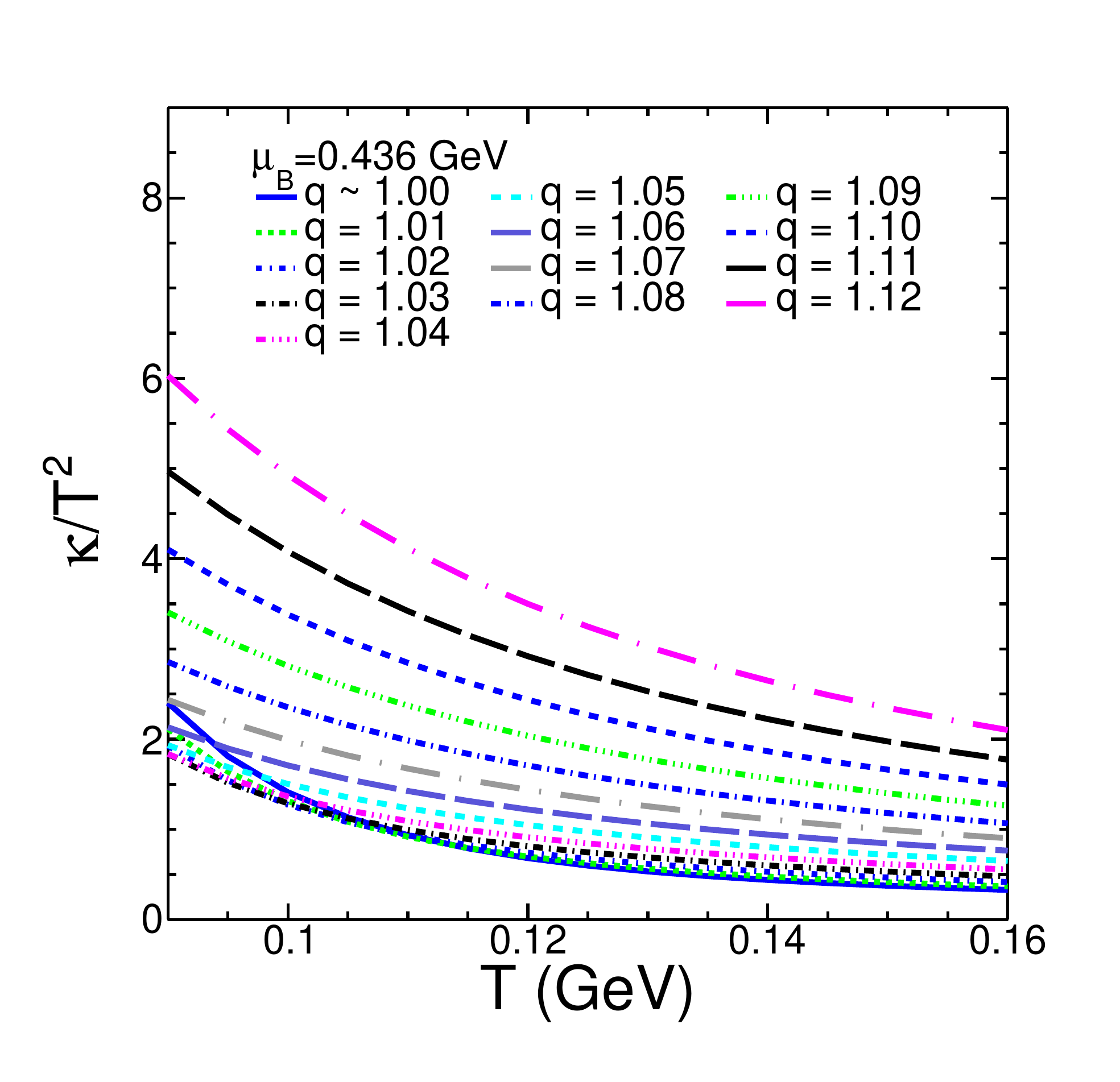}
\caption[]{(Color online) $\kappa/T^{2}$ vs $T$ plot for different $q$-values at $\mu_B=0.436$ GeV.}
\label{kappa436}
\end{figure}

In Fig.\ref{kappa25} and Fig.\ref{kappa436}, thermal conductivity $\kappa$ is shown as a function of temperature for different $q$ values for $\mu_B=25$ MeV and 436 MeV. 
Like the electrical conductivity, $\kappa$ also decreases with $T$ for all values of $q$. Also, similar to $\sigma_{el}$, the fall is sharper at lower temperature and more gradual 
at higher temperature roughly mimicking the behaviour of $\tau$. But unlike $\sigma_{el}$, $\kappa$ shows significant variation with $\mu_B$. This is expected as the non-zero contributions to $\kappa$ comes only from baryons which is sensitive to change in chemical potential. In fact, the fall of $\kappa$ with $\mu_B$ is more dramatic compared to the fall with $T$. The decrease with $\mu_B$ is also largely shaped by the decrease of $\tau$ with $\mu_B$ as well as the fact that $\kappa$ gets non-zero contribution from baryons only and $\mu_B$ essentially decides how much of the system participates in heat conduction. So contrary to $\sigma_{el}$, $\mu_B$ is the dominant parameter affecting $\kappa$.

However, the variation of $\kappa$ with $q$ shows more complicated behaviour. At large values of temperature, $\kappa$ shows clear increase with $q$. But at lower temperature, smaller values of $q$ tend to yield a larger value of $\kappa$. This behaviour has been previously observed for velocity of sound ($c_s^2$) \cite{Khuntia:2016ikm} where $c_s^2$ was higher for larger $q$ at higher temperature and lower temperature showed opposite behaviour with clear shifting of the peak towards lower temperature for higher values of $q$.

\begin{figure}[h]
\includegraphics[height=22em]{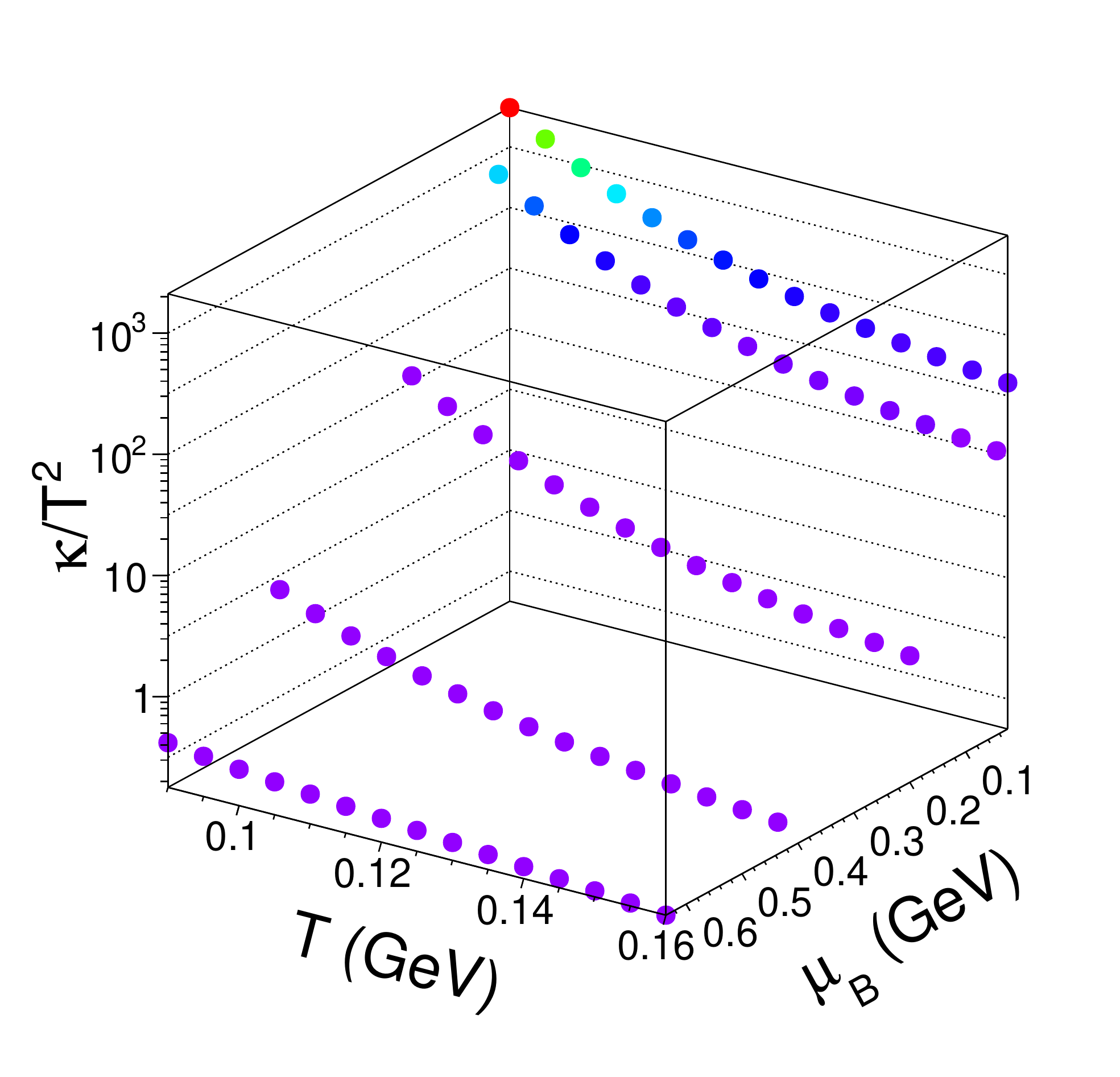}
\caption[]{(Color online)  $\kappa/T^{2}$  vs $T$ as a function of $T$ and $\mu_{B}$ for different center-of-mass energies at q $\sim 1$}
\label{kappa3D}
\end{figure}

Figure [\ref{kappa3D}] shows how $\kappa$ changes with both $T$ and $\mu_B$ for equilibrium case. As discussed, we can see that while $\kappa$ decreases with both $T$ and $\mu_B$, the change with $\mu_B$ is much sharper.

\begin{figure}[h]
\includegraphics[height=22em]{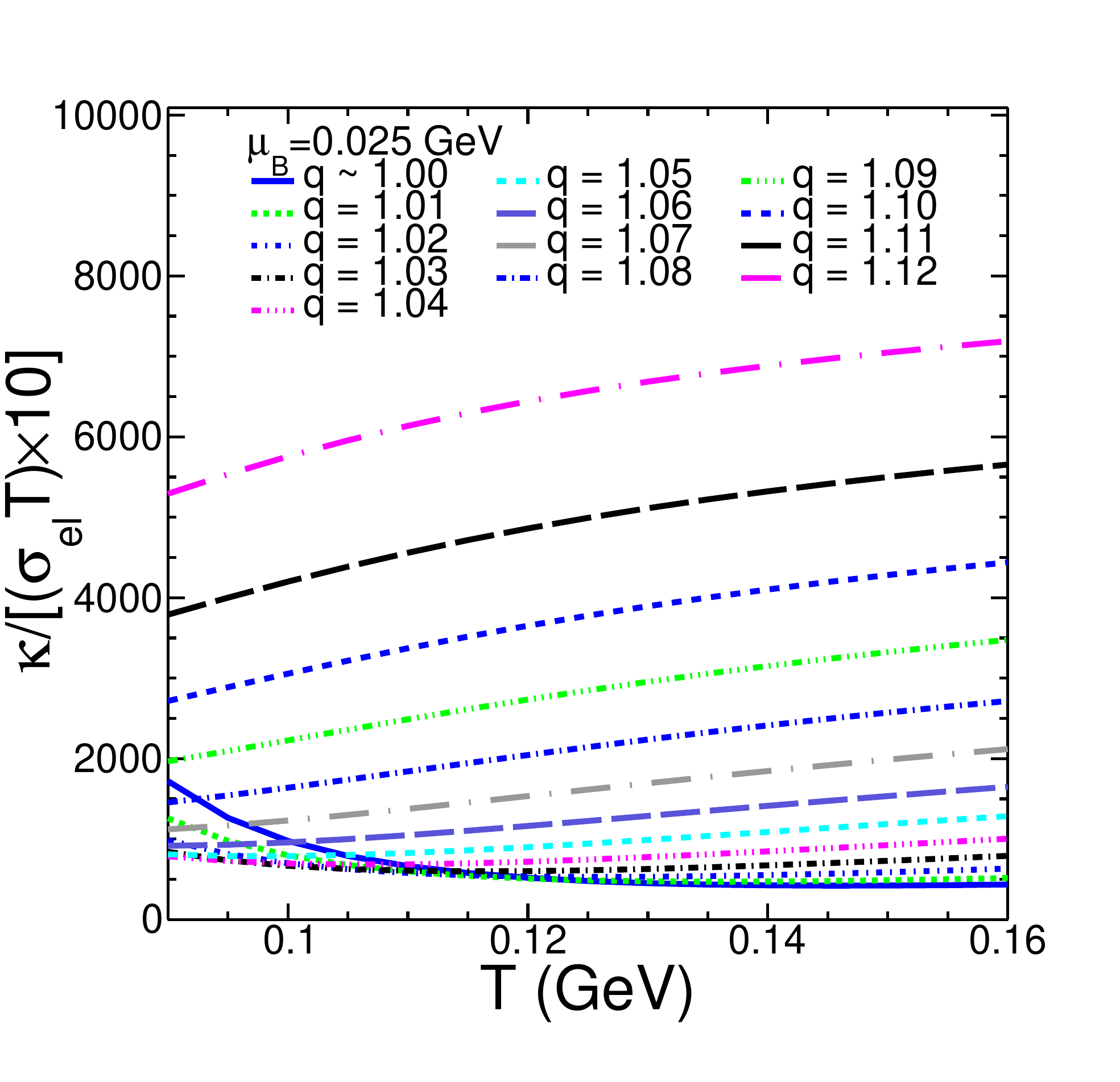}
\caption[]{(Color online) $\kappa/\sigma_{el}$ vs $T$ for different $q$-values at $\mu_B=0.025$ GeV.}
\label{wf25}
\end{figure}

\begin{figure}[h]
\includegraphics[height=20em]{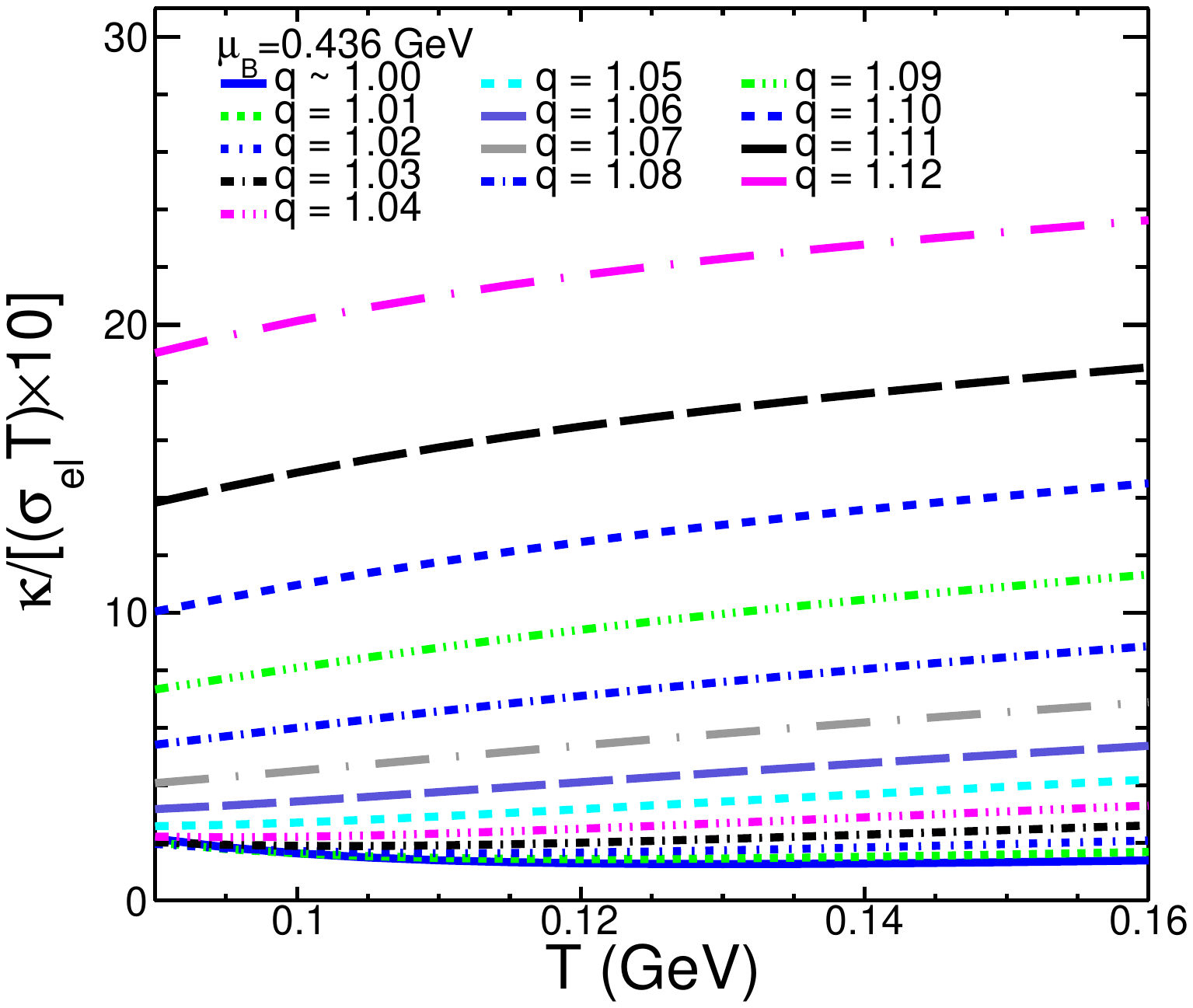}
\caption[]{(Color online)  $\kappa/(\sigma_{el}T)$  vs $T$ for different $q$-values at $\mu_B=0.436$ GeV.}
\label{wf436}
\end{figure}

Since we have calculated both $\kappa$ and $\sigma_{el}$, it would be interesting to examine the validity of Wiedemann-Franz law which states that for a material that is a good conductor of both heat and electricity, the ratio $\frac{\kappa}{\sigma_{el}}$ is proportional to temperature and the proportionality constant is called Lorenz number. In other words, $\frac{\kappa}{\sigma_{el}T}$ should remain a constant. It has been observed that metals, which is a good conductor of both electricity and heat, follow this law to a reasonably good extent. In present context, we should remember that hadron gas is very different from metals with very low electrical conduction. In Fig.[\ref{wf25}] and Fig.[\ref{wf436}], we have shown the behaviour of the Lorenz number with temperature for different $q$-values. We observed wide variation of Lorenz number with temperature. The Lorenz number decreases with temperature for equilibrium case while increasing significantly when the system is far from equilibrium. In addition, as we shall discuss in the subsequent paragraphs, beyond a certain temperature the Lorenz number tend to increase for any values of the non-extensive parameter, $q$. We shall also see that this reversal in trend happens at progressively lower temperatures as we increase the baryochemical potential. However, this violation of Wiedemann-Franz law should not come as a surprise as there are significant differences between a metal and hadron gas. In metals, at low temperature, both electrical and thermal conduction are accomplished by the same particles, i.e, the charged particles whereas in a hadronic gas, they are done by different particles. The electrical conduction is done by any charged hadrons whereas thermal conductivity requires a conserved charge which is baryon number in our case. So, $\mu_B$ affects $\kappa$ significantly while having negligible effects on $\sigma_{el}$. This ensures significant decrease in Lorenz number with changing $\mu_B$. Violation of Wiedemann-Franz law has been shown recently in the context of 2-flavor quark matter in Nambu-Jona Lasinio model \cite{Harutyunyan:2017ttz}, for hot QGP medium \cite{Mitra:2017sjo} and in quasiparticle model in presence of strong magnetic field \cite{Rath:2019vvi}.

\begin{figure}[h]
\includegraphics[height=22em]{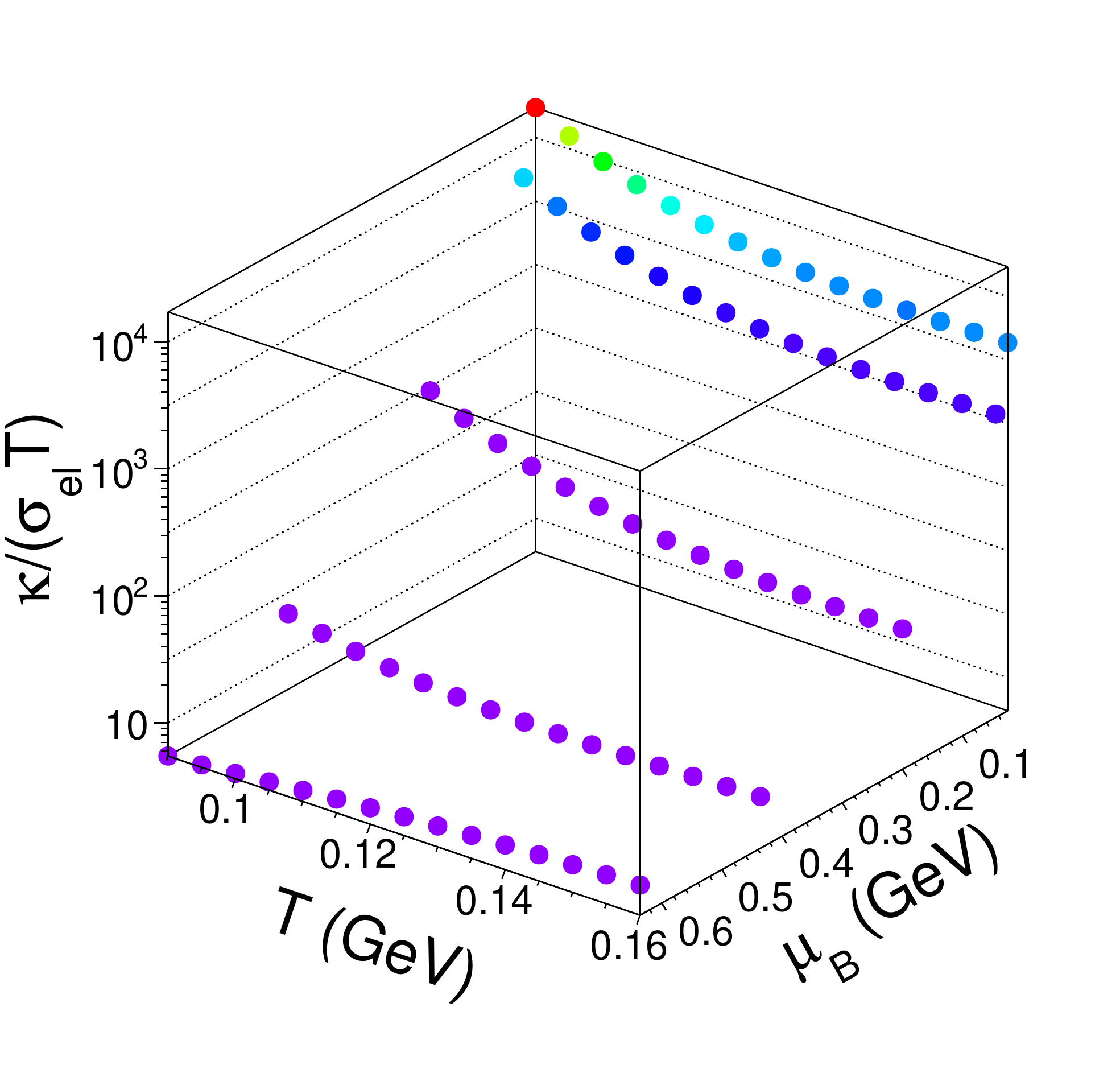}
\caption[]{(Color online)  $\kappa/(\sigma_{el}T)$  as a function of $T$ and $\mu_{B}$ for different center of mass energies at q $\sim 1$}
\label{wf3D}
\end{figure}

\begin{figure}[h]
\includegraphics[height=23em]{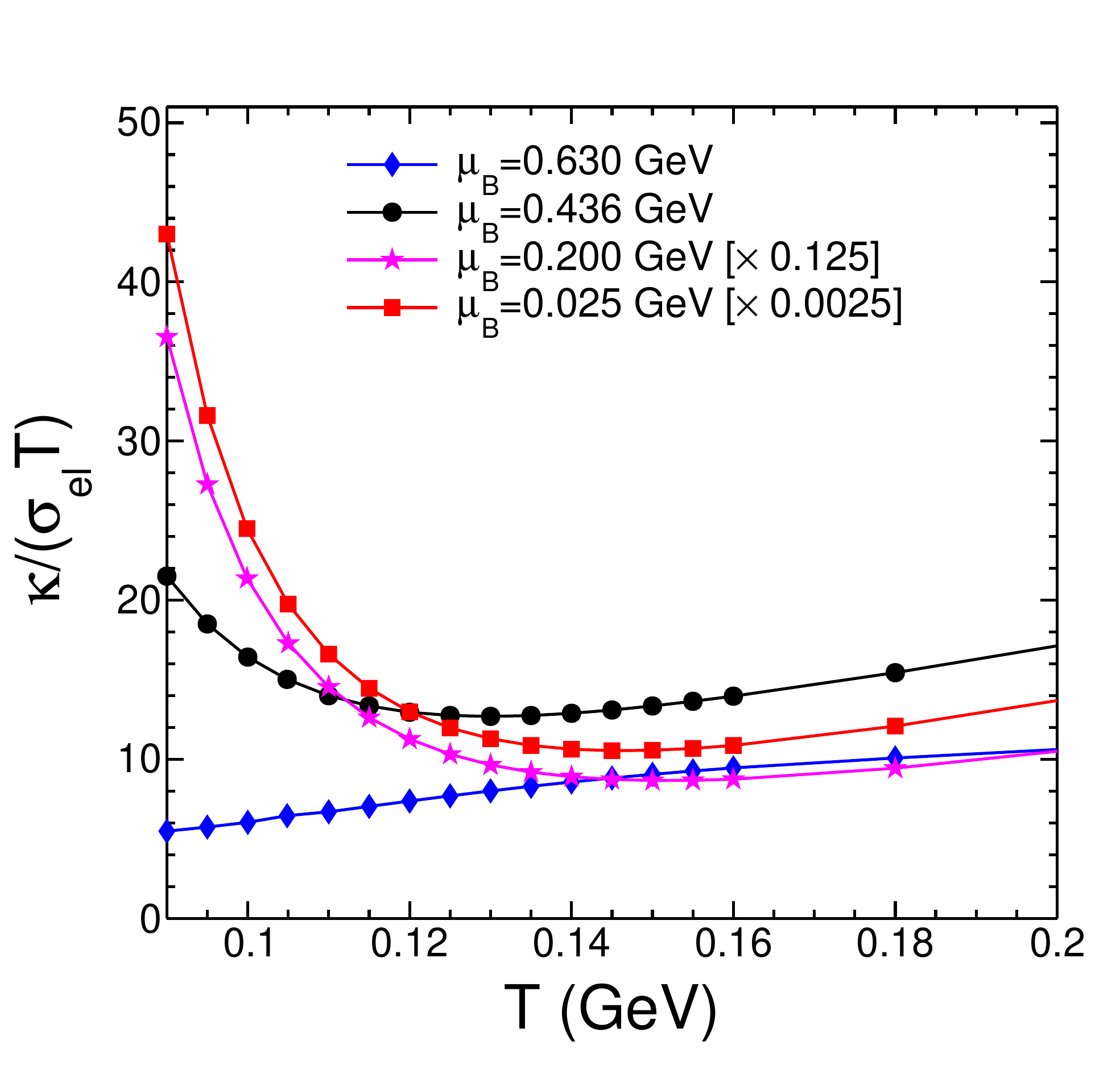}
\caption[]{(Color online)  $\kappa/(\sigma_{el}T)$, the Lorenz number for a equilibrated hot hadron gas as a function of $T$. The baryochemical potentials in legend, correspond to various collision energies discussed in the paper.}
\label{2Tc}
\end{figure}

Like $\sigma_{el}$ and $\kappa$, we have shown the variation of $L$ with both $T$ and $\mu_B$ for equilibrium scenario in Fig.[\ref{wf3D}]. The Lorenz number decreases with both $T$ and $\mu_B$ throughout the phase diagram except for $\mu_B=630$ MeV where it increases with $T$. As we can see from Fig.[\ref{kappa3D}], the change in $\kappa$ with $T$ is very slight for $\mu_B=630$ MeV and $\sigma_{el}$ changes faster thus resulting in an increasing trend for $L$. In Fig.~\ref{2Tc}, the $\kappa/(\sigma_{el}T)$, the Lorenz number is shown as a function of the temperature of a hot hadron gas at equilibrium for various values of the baryochemical potentials. It is interesting to observe a minima around $T_c$, for higher collision energies (low $\mu_B$), which slowly vanishes for a baryon rich matter expected to be formed at NICA energies for which the Lorenz number increases monotonically through out the range of temperature discussed here. This behavior is quite interesting and needs further investigation.

\begin{figure}[h]
\includegraphics[height=19em]{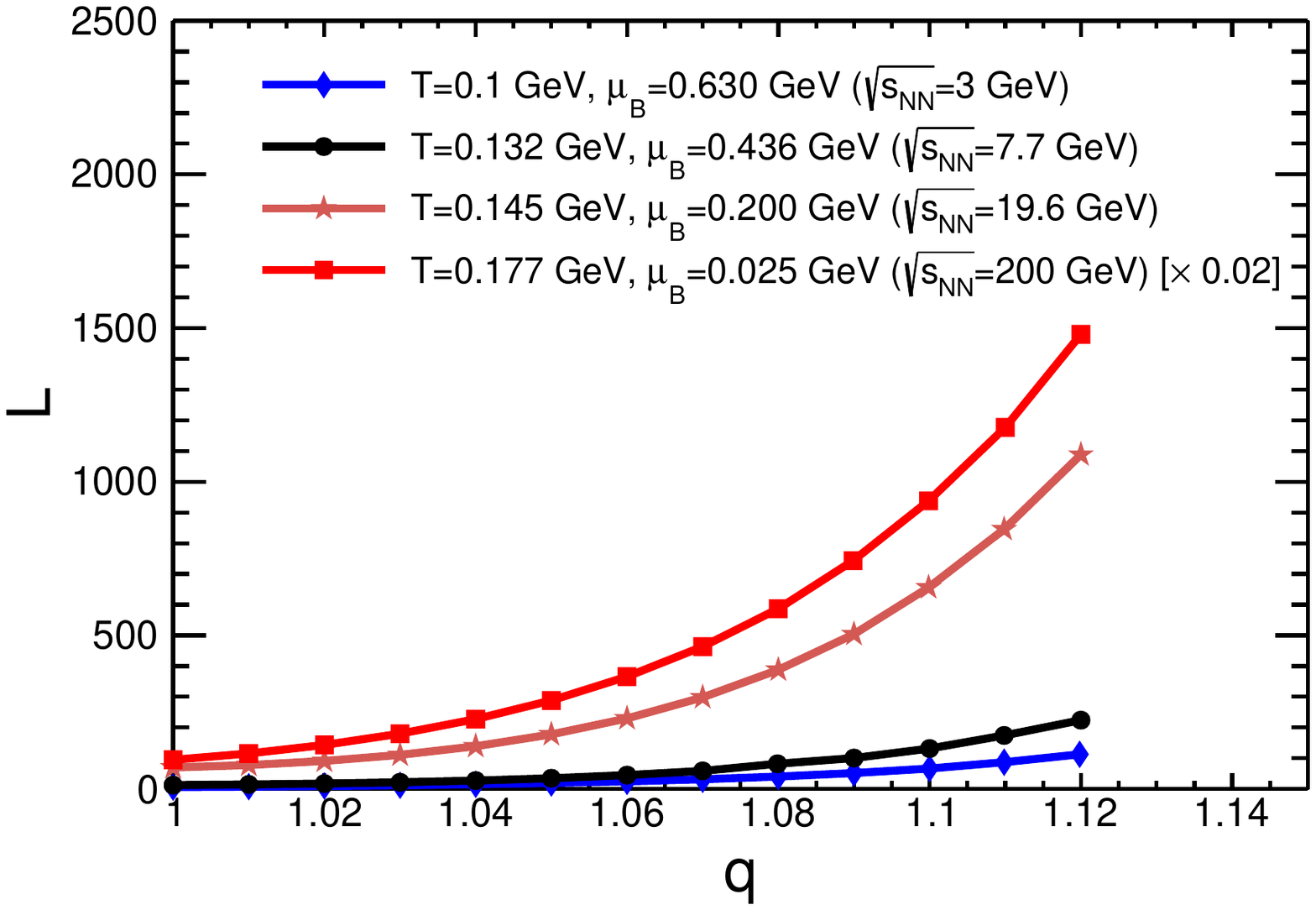}
\caption[]{(Color online)  Lorenz number as a function of the non-extensive parameter, $q$.}
\label{ln}
\end{figure}

In Fig.[\ref{ln}], we have shown the variation of the Lorenz number with the Tsallis non-extensive parameter $q$. The temperature corresponding to each $\mu_B$ in the plot indicates the chemical freezeout temperatures at the mentioned centre-of-mass energies for the relevant experiment. We see that $L$ increases slowly with $q$ for lower values of $q$ and then increases rapidly for any $\mu_B$. This happens as a result of much sharper increase in $\kappa$ with $q$ compared to $\sigma_{el}$. It is worth emphasizing again that we are considering only the freezeout temperature. Had we considered a much lower temperature, we would observe exactly opposite behaviour of $L$ with $q$ as $\kappa$ tends to decrease with increasing $q$ at lower temperature whereas the qualitative behaviour of $\sigma_{el}$ with $q$ remains the same for any temperature. At very high $\mu_B$, the Lorenz number is very small at small $q$ as $\kappa$ falls dramatically with $\mu_B$ whereas $\sigma_{el}$ remains relatively unchanged for reasons mentioned above.

\section{Summary}
\label{summary}
This work is the first attempt to investigate the non-equilibrium effects on thermal and electrical conductivities in the context of a hadron gas. The Tsallis-Boltzmann statistics has been used to include effects of non-equilibrium and solved that Boltzmann transport equation in relaxation time approximation for $\sigma_{el}$ and $\kappa$. We also checked whether the Wiedemann-Franz law holds for hadron gas.  This analysis has been carried out at five different baryon chemical potentials, $\mu_B$ which are relevant for RHIC, FAIR and NICA experiments. We can summarize our finding as follows.

a) The electrical conductivity $\sigma_{el}$ decreases as the system moves away from equilibrium, i.e, for higher $q$-values. This change is significant particularly at lower temperature. The qualitative behaviour of $\sigma_{el}$ as a function of $T$ and $\mu_B$ remains the same for both equilibrium and non-equilibrium scenario and it decreases with both $T$ and $\mu_B$ even though the change with $\mu_B$ very small.

b) The qualitative behaviour of $\kappa$ also remains same for equilibrium and non-equilibrium scenario. However, as the system moves away from equilibrium, the change in heat conduction is more complex. At lower temperature, $\kappa$ is greater for equilibrium case similar to $\sigma_{el}$. At higher temperature, particularly near the freezeout temperature, 
$\kappa$ increases significantly as the system moves farther and farther from equilibrium.

c) Wiedemann-Franz law is not obeyed in a hadron gas system. The Lorenz number, i.e, the ratio $\frac{\kappa}{\sigma_{el}T}$ decreases with temperature for equilibrium case and increases with temperature significantly as the system moves far from equilibrium. Also, for any given $T$ and $\mu_B$, the Lorenz number increases as the system moves away from equilibrium showing the dominance of heat conduction over electrical conductivity.

d) For an equilibrated hadron gas, the Lorenz number shows an interesting variation with temperature when studied for different baryochemical potentials. For higher collision energies, we observe a minima in Lorenz number around $T_c$, whereas for lower collision energy (relevant for NICA), we don't see such a behavior, rather it shows a monotonic increase with temperature.

This work by no mean presents a complete picture of hadron gas with non-equilibrium effects. First of all, the coefficients calculated here are of limited use as the macroscopic description for q-generalized system has not been developed yet. However if it is developed, then these quantities would be immensely useful, particularly in the context of pp collisions. Also, we have considered only the baryons to be eligible for heat conduction whereas it has been shown that pion number also remains almost  constant in the hadronic phase and thus making it eligible for heat conduction. So, for a more complete picture of the phase after chemical freezeout, one should include the effects of pions. Also, it would be interesting to see the non-equilibrium effects when quarks are present in the system. Although in the present work, we have considered a universal freeze-out scenario, one can also take a mass-dependent differential freeze-out scenario. Also at lower collision energies, it would be interesting to explore the present work by including van der Waal kind of interactions. Some of these works are in progress and will be reported later. 
\section*{Acknowledgements}
The authors acknowledge the financial supports  from  ALICE  Project  No. SR/MF/PS-01/2014-IITI(G) of Department of Science \& Technology, Government of India. RR and ST acknowledge the financial support by DST-INSPIRE program of Government of India. BC and ST acknowledge fruitful discussions with Dr. Guru Prakash Kadam and Prof. Hiranmaya Mishra. SKT would like to acknowledge the financial support from TEQIP-III- a joint venture of MHRD and the World Bank.

\end{document}